\documentclass[aps,prc,twocolumn,superscriptaddress]{revtex4-2}

\usepackage{graphicx}

\begin{document}

\title{TITAN mass measurements of neutron-rich Cs, Ba and $r$-process lanthanide abundances}

\author{T.-H.~Yeh}
 \affiliation{TRIUMF, 4004 Wesbrook Mall, Vancouver, British Columbia V6T 2A3, Canada}

\author{J.D. Cardona}
 \affiliation{TRIUMF, 4004 Wesbrook Mall, Vancouver, British Columbia V6T 2A3, Canada}
 \affiliation{Dept. of Physics \& Astronomy, University of Manitoba, Winnipeg, MB R3T 2N2, Canada}

\author{Y.~Wang}
 \affiliation{TRIUMF, 4004 Wesbrook Mall, Vancouver, British Columbia V6T 2A3, Canada}
 \affiliation{Department of Physics and Astronomy, University of British Columbia, Vancouver, British Columbia, V6T 1Z1, Canada}

\author{J. Ash}
 \affiliation{TRIUMF, 4004 Wesbrook Mall, Vancouver, British Columbia V6T 2A3, Canada}

\author{B. Ashrafkhani}
 \affiliation{TRIUMF, 4004 Wesbrook Mall, Vancouver, British Columbia V6T 2A3, Canada}
 \affiliation{Department of Physics and Astronomy, University of Calgary, Calgary, AB, T2N 1N4, Canada}

\author{I. Belosovic}
 \affiliation{TRIUMF, 4004 Wesbrook Mall, Vancouver, British Columbia V6T 2A3, Canada}

\author{J. Bergmann}
 \affiliation{Physikalisches Institut, Justus-Liebig-Universit\"{a}t Gie\ss en, 35392 Gie\ss en, Germany}

\author{E. Dunling}
 \affiliation{TRIUMF, 4004 Wesbrook Mall, Vancouver, British Columbia V6T 2A3, Canada}

\author{L. Egoriti}
 \affiliation{TRIUMF, 4004 Wesbrook Mall, Vancouver, British Columbia V6T 2A3, Canada}
 \affiliation{Department of Physics and Astronomy, University of British Columbia, Vancouver, British Columbia, V6T 1Z1, Canada}
 
\author{G. Gelinas}
 \affiliation{TRIUMF, 4004 Wesbrook Mall, Vancouver, British Columbia V6T 2A3, Canada}
 \affiliation{Department of Physics and Astronomy, University of Calgary, Calgary, AB, T2N 1N4, Canada}

\author{G. Gwinner}
\affiliation{Dept. of Physics \& Astronomy, University of Manitoba, Winnipeg, MB R3T 2N2, Canada}

\author{Z. Hockenbery}
 \affiliation{TRIUMF, 4004 Wesbrook Mall, Vancouver, British Columbia V6T 2A3, Canada}

\author{C. Izzo}
 \affiliation{TRIUMF, 4004 Wesbrook Mall, Vancouver, British Columbia V6T 2A3, Canada}

\author{A. Jacobs}
 \affiliation{TRIUMF, 4004 Wesbrook Mall, Vancouver, British Columbia V6T 2A3, Canada}
 \affiliation{Department of Physics and Astronomy, University of British Columbia, Vancouver, British Columbia, V6T 1Z1, Canada}

\author{S. Kakkar}
 \affiliation{TRIUMF, 4004 Wesbrook Mall, Vancouver, British Columbia V6T 2A3, Canada}
 \affiliation{Dept. of Physics \& Astronomy, University of Manitoba, Winnipeg, MB R3T 2N2, Canada}

\author{B. Kootte}
 \affiliation{TRIUMF, 4004 Wesbrook Mall, Vancouver, British Columbia V6T 2A3, Canada}
 \affiliation{Dept. of Physics \& Astronomy, University of Manitoba, Winnipeg, MB R3T 2N2, Canada}

\author{E. M. Lykiardopoulou}
 \affiliation{TRIUMF, 4004 Wesbrook Mall, Vancouver, British Columbia V6T 2A3, Canada}
 \affiliation{Department of Physics and Astronomy, University of British Columbia, Vancouver, British Columbia, V6T 1Z1, Canada}

\author{T. Murb{\"o}ck}
 \affiliation{TRIUMF, 4004 Wesbrook Mall, Vancouver, British Columbia V6T 2A3, Canada}

\author{A. Mollaebrahimi}
 \affiliation{TRIUMF, 4004 Wesbrook Mall, Vancouver, British Columbia V6T 2A3, Canada}
 \affiliation{Physikalisches Institut, Justus-Liebig-Universit\"{a}t Gie\ss en, 35392 Gie\ss en, Germany}
 \affiliation{GSI Helmholtzzentrum für Schwerionenforschung GmbH, 64291 Darmstadt, Germany}

\author{A. Ridley}
 \affiliation{TRIUMF, 4004 Wesbrook Mall, Vancouver, British Columbia V6T 2A3, Canada}
 \affiliation{Dept. of Physics, University of Surrey, Surrey, UK}

\author{S. F. Paul}
 \affiliation{TRIUMF, 4004 Wesbrook Mall, Vancouver, British Columbia V6T 2A3, Canada}

\author{W. S. Porter}
 \affiliation{TRIUMF, 4004 Wesbrook Mall, Vancouver, British Columbia V6T 2A3, Canada}
 \affiliation{Department of Physics and Astronomy, University of British Columbia, Vancouver, British Columbia, V6T 1Z1, Canada}

\author{M. P. Reiter}
 \affiliation{School of Physics and Astronomy,University of Edinburgh, Edinburgh, EH9 3FD, United Kingdom}

\author{J. Ringuette}
 \affiliation{TRIUMF, 4004 Wesbrook Mall, Vancouver, British Columbia V6T 2A3, Canada}
 \affiliation{Dept. of Physics and Astronomy, Golden, Colorado, USA}

\author{R. Simpson}
 \affiliation{TRIUMF, 4004 Wesbrook Mall, Vancouver, British Columbia V6T 2A3, Canada}
 \affiliation{Department of Physics and Astronomy, University of British Columbia, Vancouver, British Columbia, V6T 1Z1, Canada}

 \author{C. Walls}
 \affiliation{TRIUMF, 4004 Wesbrook Mall, Vancouver, British Columbia V6T 2A3, Canada}
 \affiliation{Dept. of Physics \& Astronomy, University of Manitoba, Winnipeg, MB R3T 2N2, Canada}

\author{M.~R.~Mumpower}
 \affiliation{Department of Physics, University of Notre Dame, Notre Dame, Indiana 46556, USA}
 \affiliation{Center for Theoretical Astrophysics, Los Alamos National Laboratory, Los Alamos, New Mexico 87545, USA}

\author{N.~Vassh}
\email{Contact author: nvassh@triumf.ca}
 \affiliation{TRIUMF, 4004 Wesbrook Mall, Vancouver, British Columbia V6T 2A3, Canada}

\author{A.A. Kwiatkowski}
 \affiliation{TRIUMF, 4004 Wesbrook Mall, Vancouver, British Columbia V6T 2A3, Canada}
 \affiliation{Dept. of Physics \& Astronomy, University of Victoria, Victoria, BC V8P 5C2, Canada}

\date{\today}

\begin{abstract}
We present measurements for the masses of five neutron-rich isotopes, $^{149-151}$Cs and $^{151, 152}$Ba, probed for the first time by TITAN at TRIUMF with time-of-flight measurement techniques. We propagate these masses to the nuclear reaction and decay data required for the simulation of the rapid neutron capture process (r-process) nucleosynthesis in neutron star mergers. We show that these neutron-rich masses affect the abundance predictions near mass number $A\sim148-152$ corresponding to lanthanide element abundances at $Z=60,\,62$ and $63$. We demonstrate that these new TITAN masses smooth out the odd-even effect in isotopic abundance predictions near $A\sim150$ in both fission cycling astrophysical conditions and conditions that do not reach actinides. We further show that these new masses adjust how fission fragments settle into place when forming the final abundances, and consider the effect on comparisons with stellar abundance ratios such as [Ag/Eu], [Sm/Eu], and [Nd/Eu].
\end{abstract}


\maketitle

\section{Introduction}
Abundance patterns of elements and isotopes in stars and our Sun have been used to decipher the ultimate origin of elements for decades \cite{B2FH,Arcones:2016euo,Schatz:2022vzq,Horowitz:2018ndv,Kajino:2019abv,Cowan:2019pkx}. In particular, lanthanide elements display interesting stellar abundance trends that have been argued to point to fission cycling as a source of the so-called `robustness' of lanthanide abundances \cite{Beun:2007wf,Sneden2008,Goriely:2011vg,Korobkin:2012uy,deJesusMendoza-Temis:2014owk,Goriely:2015oha}. Additionally, correlations between lanthanides like Eu and Ag provide hints that the two elements can be co-produced \cite{Vassh:2019cey,Roederer:2023spd}. Lanthanide properties and abundances also play a key role in conclusions that the electromagnetic counterpart of the GW170817 neutron star merger pointed to the production of heavy elements in this astrophysical event \cite{Metzger:2019zeh,Radice:2020ddv,Margutti:2020xbo,Zhu:2020eyk,Barnes:2020nfi}. Thus, lanthanide abundances and abundance ratios serve an illuminating role, and they have notable ties to traces of fission in astrophysical scenarios. 

With the prospect of refining our interpretation of lanthanide abundances, we consider rapid neutron capture ($r$-process) nucleosynthesis in neutron star mergers and explore the details of how the abundances of key lanthanide species such as Eu and Nd are finalized. Although many distinct types of ejecta are possible from neutron star mergers, one defining characteristic is whether the conditions present in the ejecta host fission. If the ejecta conditions support production of actinides (e.g. conditions are neutron rich enough for neutron captures to reach nuclei with $Z>88$), then fission fragments can be populated. These fission fragments then have an imprint on the final abundance pattern of the ejecta from an $r$-process event, which can then be observed later on the surface of stars that bear the mark of events that enriched their gasses. If actinide elements cannot be produced (e.g. conditions are insufficiently neutron rich), then the relative abundances of the lanthanides are influenced by mass trends that are more local (i.e. near similar mass numbers), as well as the properties of lighter species such as near $N$=82 that are the seeds that can ultimately go on to populate the lanthanides. Therefore, experimental measurements can impact the interpretation of Solar and stellar abundances by refining the nuclear data impacting lanthanide production in the $r$ process.

To provide such constraints on lanthanide abundances, we use the multiple-reflection time-of-flight mass spectrometer (MR-TOF-MS) at TITAN at TRIUMF to perform mass measurements of $^{149-151}$Cs ($Z=55$, $N=94, 95, 96$) and $^{151,152}$Ba ($Z=56$, $N=95, 96$) isotopes. These measurements push the bounds of experimentally probed species into a presently desolate region of the nuclear chart where isotopes with $N>95$ have previously not been reached below element number $Z<58$. 

We present the mass values of these newly probed species as well as the experimental method used to determine these masses in Sec.~\ref{sec:mass}. In Sec.~\ref{sec:s2nrxndecay} we consider the nuclear structure implications and propagate these new mass measurements to the neutron capture, $\beta$-decay and neutron separation energies that enter $r$-process calculations. In Sec.~\ref{sec:rprocess}, we show results for $r$-process abundance calculations both in cases that host fission and in cases that do not in order to evaluate how Cs and Ba nuclear properties influence the freshly deposited fission fragments when forming the final $r$-process abundances. We conclude in Sec.~\ref{sec:conclude}.

\section{Mass measurements with TITAN}\label{sec:mass}

The cesium and barium isotopes measured during this experiment were produced at the TRIUMF Isotope Separator and Accelerator facility (ISAC) \cite{JensISAC} during the initial use of the new proton to neutron converter target \cite{lucaeg}. An 80 $\mu$A proton beam at an energy of 480 MeV impinged onto a p-to-n target to produce rare nuclei via neutron-induced fission on natural uranium. Through surface ionization Cs and Ba were produced. After the neutron-induced fission products leave the target, the radioactive ion beam (RIB) is filtered to a single mass unit by the ISAC magnetic mass separator with a resolving power of $m/ \Delta m \approx 2000$ \cite{JensISAC}.

The 20 keV beam was transported to TITAN's helium-gas-filled radiofrequency quadrupole (RFQ) cooler-buncher \cite{rfq2}. Bunched ions are then pulsed down to an energy of 1.3 keV at a rate of 50 Hz and guided to the MR-TOF-MS \cite{reiter21,Jesch2015-iu} for the mass measurement. The MR-TOF-MS includes an injection trap, where ions are re-cooled, and the time-of-flight (TOF) analyzer: a pair of opposing ion mirrors separated by a drift tube. This measurement was performed using the mass-selective re-trapping mode \cite{Dickel,Jacobs_2019} to enhance the relative abundance of the ion of interest. In this re-trapping mode the beam is stored between the mirrors, for a defined number of isochronous turns (IT) until there is enough separation between the ions of interest (IOI) and isobaric contaminants in time-of-flight. For this work ions performed N = 650-680 IT. From there the IOI is re-trapped and cooled in the MR-TOF RFQ, and the contaminants are deflected out of the analyzer by the deflecting electrodes. This re-trapping mode enables the suppression of isobaric contamination by up to 4 orders of magnitude as seen in previous TITAN experiments\cite{beck21,Mukul_21,Izzo_21,Porter_22}. Finally, the beam is re-injected into the analyzer for the mass measurement. The beam undergoes more IT and is passed to the MagneTOF detector (ETP MagneTOF$^{\rm TM}$) for TOF recording. 

\begin{figure}[!ht]
\centering
\includegraphics[scale=0.14]{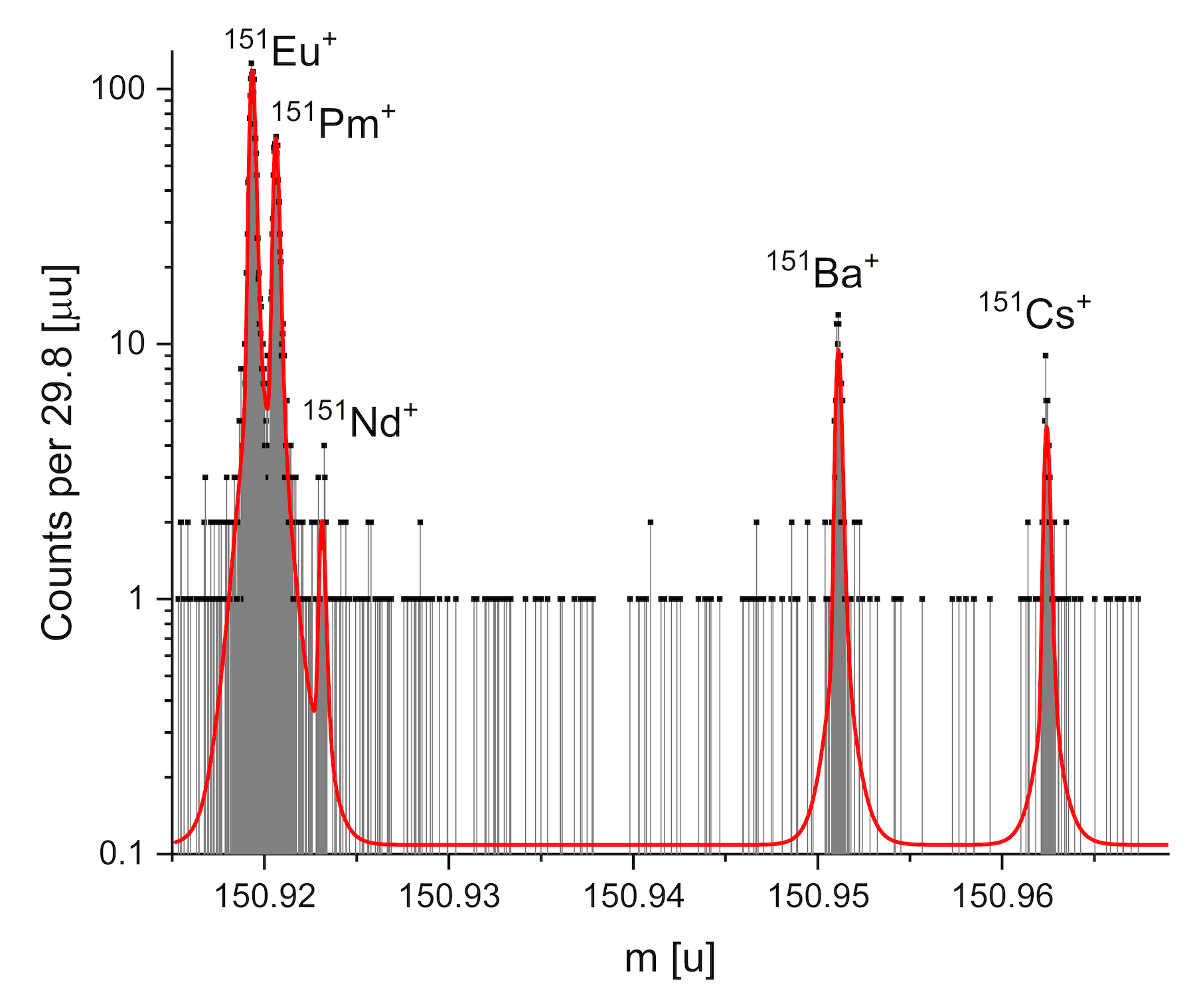}\\
\caption{TOF spectrum for $A$=151, with the TOF converted into mass $m$. So-called re-trapping \cite{beck21} reduced the lanthanide contamination to reveal fully resolved Cs and Ba peaks. With the hyper-EMG fit \cite{EMG} to the data in red. } 
\label{fig:massmeas}
\end{figure}

\begin{table}[!h]
\centering
{\footnotesize	   
    \begin{tabular}[c]{|c|c|c|c|c|}\hline
 Nuclide& Mass Calibrant& TITAN& AME2020& TITAN$-$AME\\\hline \hline 
         $^{149}$Cs& $^{149}$Sm & -43406(21)&  -43300\#(400\#)&  -105(401)\\ \hline
$^{149}$Ba& $^{149}$Sm& -52809(42)&  -52831(2.5)&  22(42)\\ \hline
         $^{150}$Cs& $^{150}$Sm& -38255(23)&  -38170\#(400\#)&  -85(401)\\ \hline
         $^{150}$Ba& $^{150}$Sm& -49893(29)&  -49890(6)&  -3(30)\\ \hline 
         $^{151}$Cs& $^{151}$Pm &-34460(26)&  -34280\#(500\#)&  -180(501)\\ \hline 
         $^{151}$Ba& $^{151}$Pm &-44982(22)&  -44940\#(400\#)&  -42(401)\\ \hline 
         $^{152}$Ba& $^{152}$Sm& -41717(38)&  -41610\#(400\#)&  -107(402)\\\hline
    \end{tabular}}
    \caption{Atomic Mass Excess Values (keV) obtained in this work compared to extrapolated (\#) and known values from AME2020 \cite{AME2020}. Calibrants listed in the table were measured simultaneous to the Cs and Ba.}
    \label{tab:masses}
\end{table}

Masses were then determined through data analysis process laid out in \cite{AYET, SFPaul}. A time-resolved calibration (TRC) \cite{AYET} is performed for each TOF spectrum to correct for drifts in the peaks caused by temperature-induced shifts in the mirror power supplies increasing the resolving power up to ~300000 for a 33 ms measurement cycle time. Using the emgfit Python package \cite{Paul3} the spectra were fitted with hyper-exponentially modified Gaussian (hyper-EMG) line shapes \cite{EMG}. Fig.~\ref{fig:massmeas} is an example of mass $A$ = 151 after the data analysis process; similar spectra were obtained for the masses listed in Table \ref{tab:masses}. All ions measured in $1^+$. Statistical and systematic uncertainties were determined using procedures also described in \cite{AYET, SFPaul}, with a systematic uncertainty of $\delta m/m = 1\times10^{-7}$ used for these measurements. With the addition of the statistical and systematic uncertainties resulting in absolute uncertainties $\delta m < 50$ keV.     

\begin{figure*}[t]
\centering
\includegraphics[width = 0.7\textwidth]{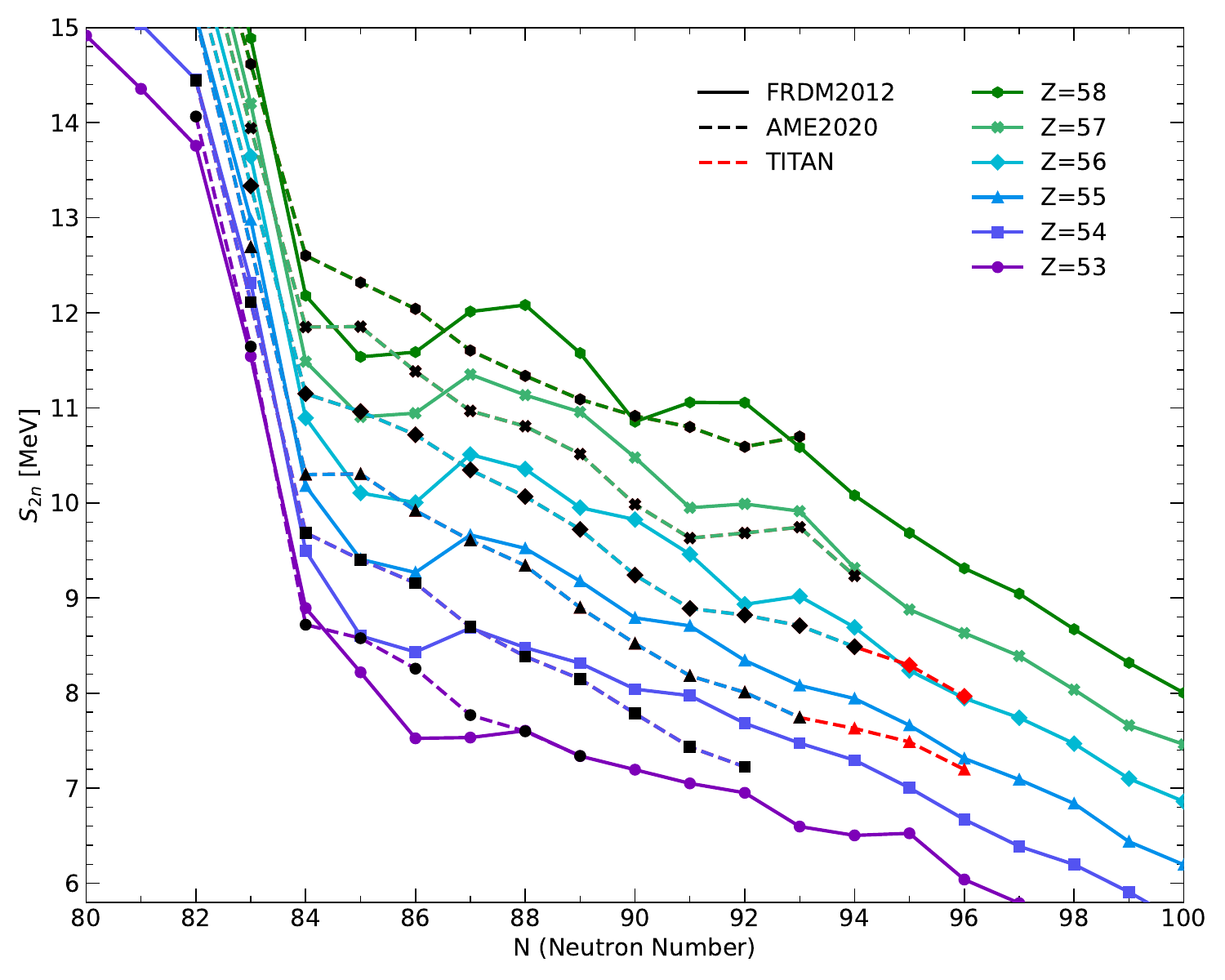}\\
\caption{Trends in the two-neutron separation energies ($S_{2n}$) of Z = 53 (I), 54 (Xe), 55 (Cs), 56 (Ba), 57 (La), and 58 (Ce) isotopic chains. FRDM2012 predictions are shown by color-coded solid lines with different markers for each $Z$. AME2020 experimental $S_{2n}$ are marked by dashed lines with black symbols, while the red ones indicate the TITAN $S_{2n}$ extension beyond AME2020.}
\label{fig:s2n}
\end{figure*}

\section{Implications for Nuclear Structure and Reaction / Decay Rates}\label{sec:s2nrxndecay}

\begin{figure}[h]
\centering
\includegraphics[width = 0.47\textwidth]{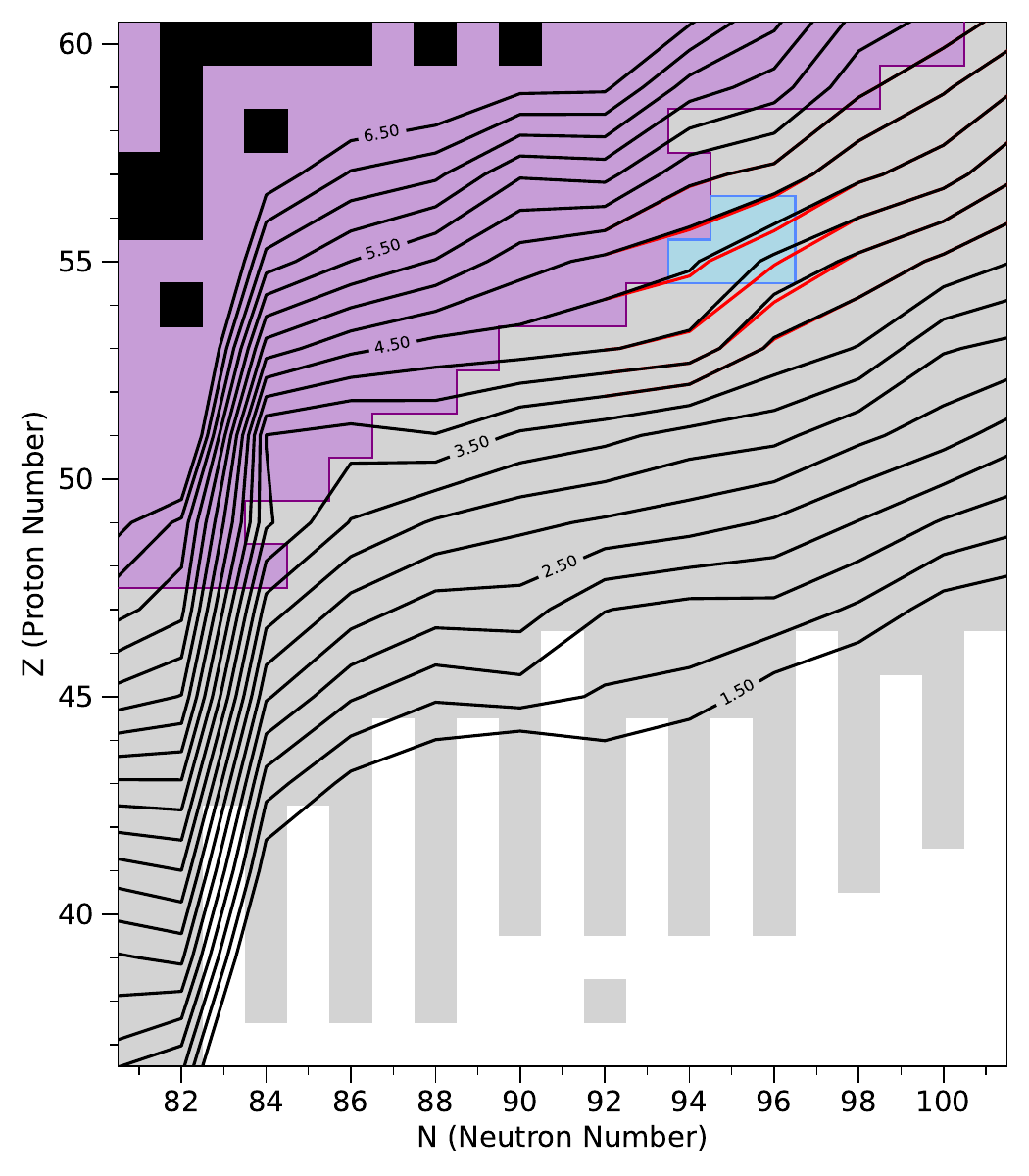}\\
\caption{Contours of constant one-neutron separation energy $S_n$ (determined here for odd $Z$ and even $N$ species) for FRDM2012+AME2020 (black) with the impact of TITAN masses on $S_n$ shown by the red contour lines. The spacing between contours is 0.25 MeV. Solid black boxes are the stable species, and the purple region indicates nuclei with known masses. Species probed by TITAN in this work are colored in light blue, and the predicted FRDM2012 dripline is shown by the lower edge of the light grey region.}
\label{fig:NZSn}
\end{figure}

We examine how these masses inform the picture of local nuclear structure via the two-neutron separation energy ($S_{2n}$) trends for $Z$=53-58 which includes the $Z$=55,56 isotopic chains of Cs, Ba respectively. Here,
\begin{equation}
    S_{2n} = M(Z,N-2) - M (Z,N) + 2M(n) 
\end{equation} 
Fig.~\ref{fig:s2n} shows that along these chains the FRDM2012 \cite{FRDM2012} and AME2020 \cite{AME2020} predictions have some significant differences. However, the most neutron-rich AME points at $N=93,94$ approach FRDM2012 predictions. TITAN measurements reaching into $N$=95,96 also move experimentally determined separation energy trends towards FRDM predictions for Cs, Ba chains. Nevertheless, overall the experimental data, including the latest TITAN measurements, show smoother trends than predicted by the FRDM2012 model. This can also be seen by examining the contours of constant one-neutron separation energy ($S_n$) as in Fig.~\ref{fig:NZSn} (here determined by odd $Z$, even $N$ isotopes in order to highlight TITAN Cs measurements). These contours are of significance to nucleosynthesis calculations, since during (n,$\gamma$) $\leftrightarrow$ ($\gamma$, n) equilibrium the $r$-process path lies along contours of constant one-neutron separation energy. This is one way in which relative abundances are affected by local structure (trends in separation energy) determined by masses along the $r$-process path.

\begin{figure}[h]
\centering
\includegraphics[width = 0.5\textwidth]{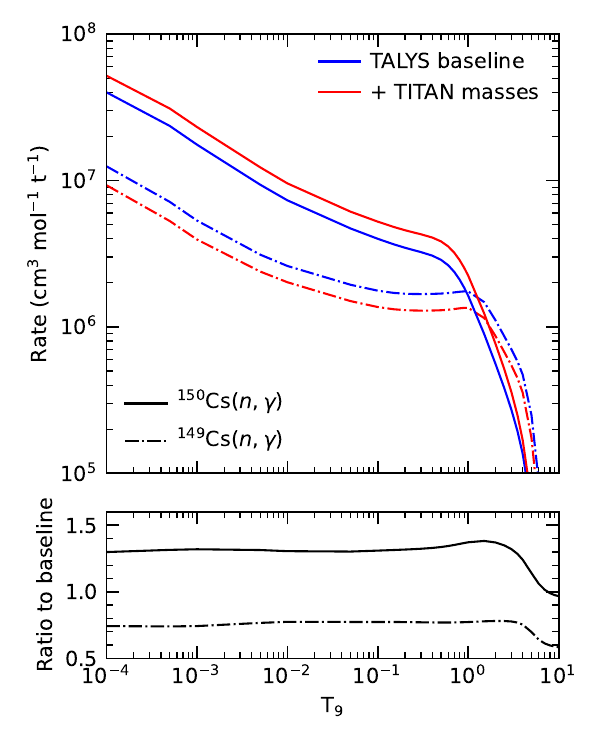}
\caption{Thermonuclear neutron capture rates of $^{149}$Cs (dash-dotted curves) and $^{150}$Cs (solid curves). The blue curves represent the baseline rates calculated from TALYS with FRDM2012 theory masses. The red curves are the updated rates with TITAN masses. The bottom panel shows the ratios of our new rates to their corresponding baselines.}
\label{fig:ncap}
\end{figure}

Another manner in which masses influence the nuclear data of relevance for the $r$-process is through the dependence of neutron capture rates on separation energies. In order to update also neutron-capture rates to reflect consistently the TITAN masses utilized for $S_n$, we utilize TALYS-2.0 \cite{TALYS}. As an example of TITAN's impact shown in Fig.~\ref{fig:ncap}, adding TITAN masses into TALYS calculations lowers the predicted neutron capture rate of $^{149}$Cs (even $N$), but increases the predicted capture rate of $^{150}$Cs (odd $N$) ( both by $\sim25\%$). 

In addition to neutron capture rates, $\beta$-decay rates also play an important role in shaping final lanthanide abundances. For the $^{149-151}$Cs and $^{151,152}$Ba isotopes considered in this work, all $\beta$-decay half-lives have been measured and are reported in NUBASE2020 decay data \cite{NUBASE2020}. So, the new masses will not enter the decay rate considered in nucleosynthesis calculations. However, $P_n$ values (neutron emission probabilities) have yet to be determined for all these species. Here we use the BeOH code from Los Alamos National Laboratory \cite{MumpowerBeoH} to consider theoretical $P_n$ values determined using TITAN masses. Given AME2020 masses along with those presented in this work, only theoretical $P_n$ value updates for $^{150,151}$Cs and $^{151}$Ba are possible since experimental $P_n$ values for $^{149}$Cs have already been adopted in our NUBASE2020 decay dataset. Note that although the $P_n$ values for $^{151}$Ba have been recently reported by RIKEN \cite{BRIKEN}, here we only consider theoretical updates for the $P_n$ value assumptions in our NUBASE2020 dataset. Our baseline dataset utilizes the commonly adopted $P_n$ values of Moller et al. 2003 \cite{MollerPn}, but replaced with experimentally determined $P_n$ when available. For the $^{150,151}$Cs nuclei we consider here, the Moller predicted values are $P_{0n}=0.56, 0.206$, $P_{1n}=0.44, 0.789$, and $P_{2n}=0.00, 0.005$ respectively, and when BeOH predictions take TITAN masses into account we get $P_{0n}=0.10, 0.15$, $P_{1n}=0.89, 0.85$, and $P_{2n}=0.01, 0.00$ respectively. For $^{151}$Ba, Moller predictions are $P_{0n}=0.957$, $P_{1n}=0.043$, and $P_{2n}=0.00$ and calculations which consider TITAN masses give $P_{0n}=0.94$, $P_{1n}=0.06$, and $P_{2n}=0.00$. And so $P_n$ updates from considering TITAN masses are modest for $^{151}$Cs and $^{151}$Ba, but in the case of $^{150}$Cs the probability of $\beta$-delayed neutron emission doubles.

\section{Impact on $r$-process abundance predictions}\label{sec:rprocess}

For the nucleosynthesis calculations we use the PRISM-1.6.0 network \cite{Sprouse:2020lpv}, where the underlying theoretical mass model is taken to be FRDM2012, fission barriers are from FRLDM model predictions and fission yields are that of GEF2016 (as in \cite{VasshJPhysG}). The theoretical $\beta$-decays and $\beta$-delayed fission rates implemented also make use of FRLDM fission barriers as in \cite{Mumpowerbdf}. For experimental decay rates, we use NUBASE2020 half-lives and branching ratios where available. For neutron capture rates we use TALYS-2.0 first with FRDM2012 and AME2020 masses, then with TITAN masses included in the experimental dataset (as in Fig.~\ref{fig:ncap}). 

\begin{figure*}[ht]
\centering
\includegraphics[width = 0.925\textwidth]{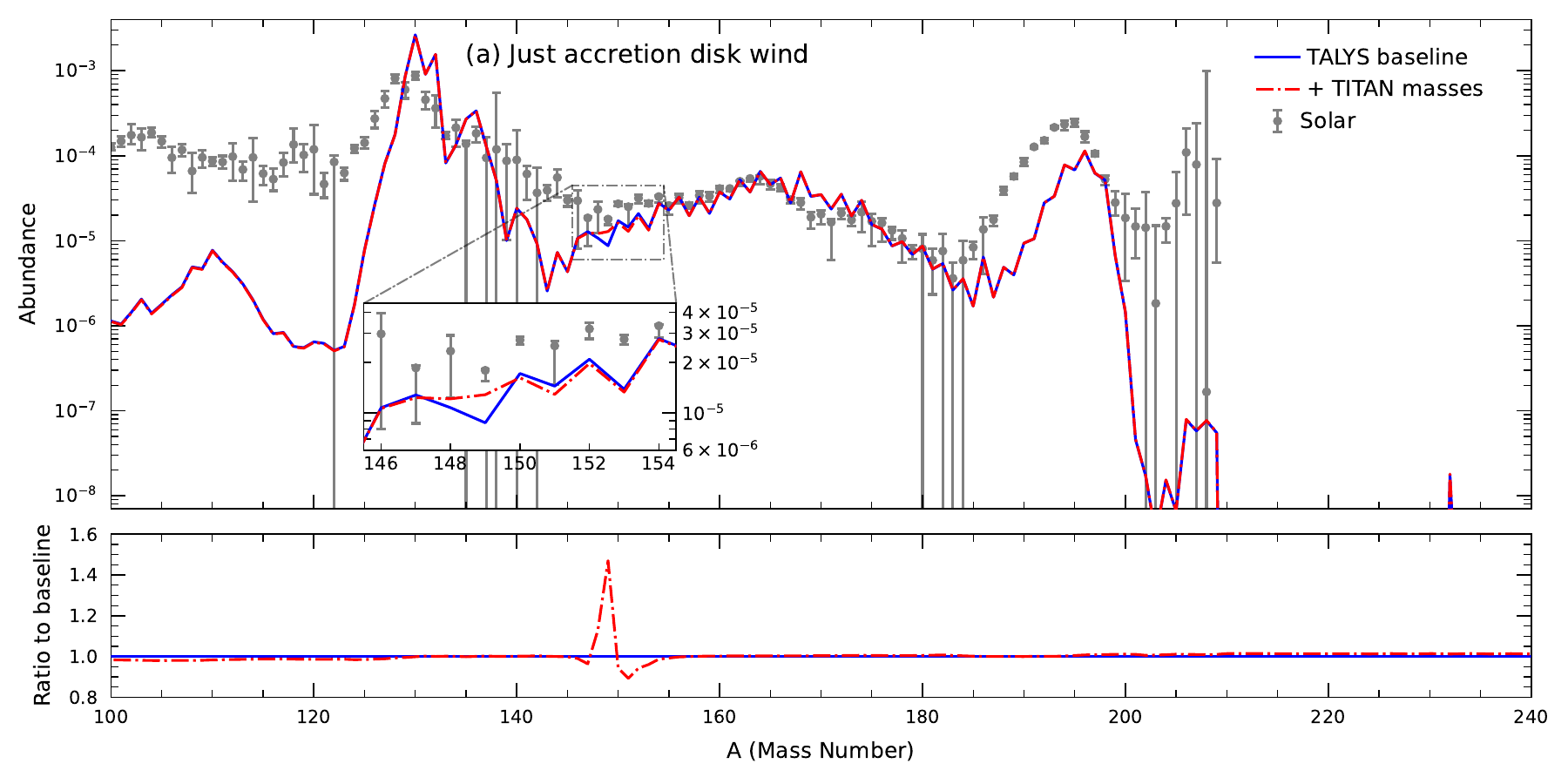}
\hspace{0.25cm}
\includegraphics[width = 0.925\textwidth]{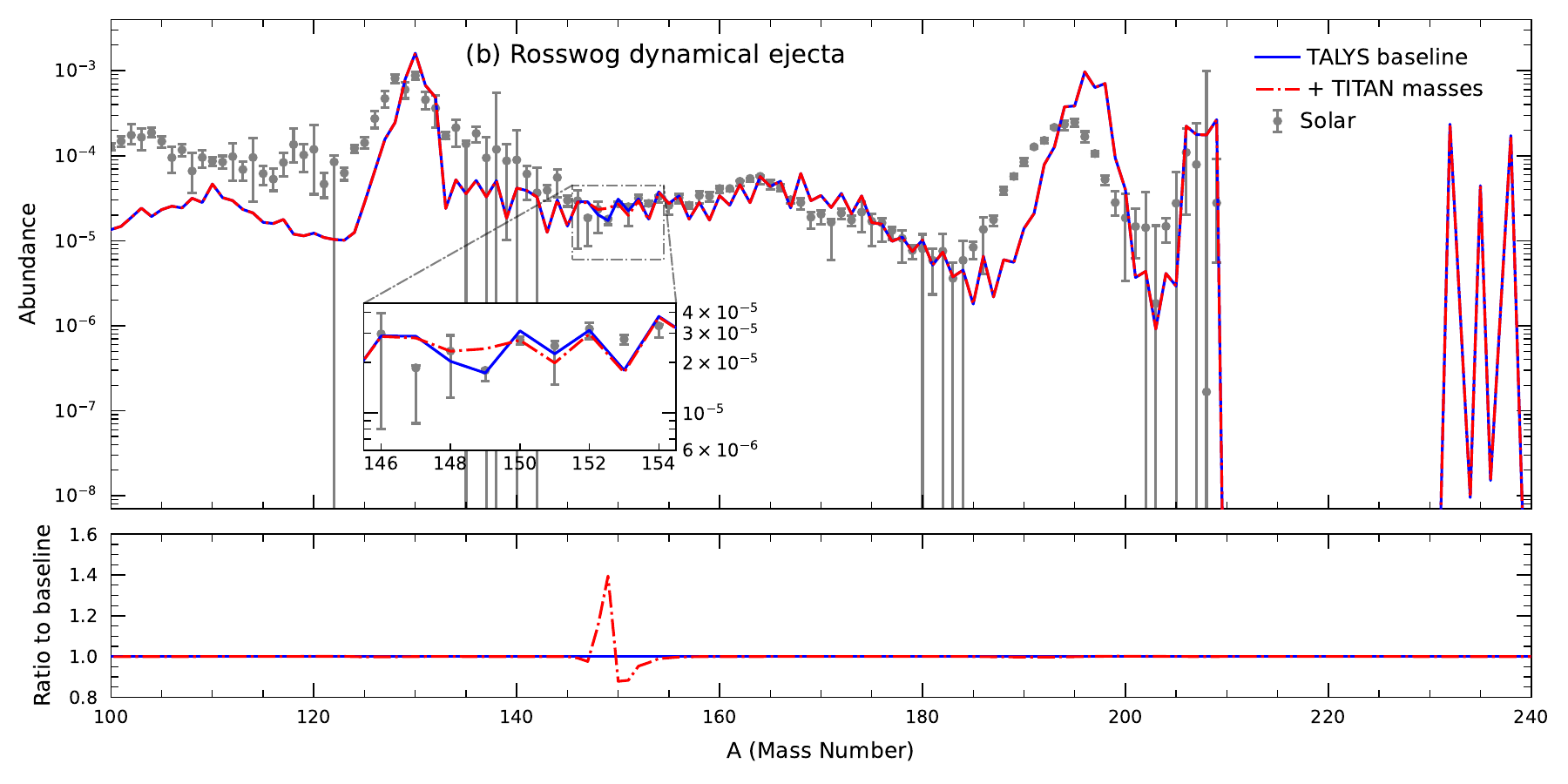}
\caption{Abundance predictions for (a) neutron star merger ejecta that does not undergo fission cycling (top) \cite{Just}, as compared to (b) very neutron-rich ejecta that does (bottom) \cite{Rosswog}. The blue solid line is the prediction of the baseline case using TALYS+FRDM2012, while the red dash-dotted line shows the impact of TITAN masses. Abundances have been rescaled to Solar abundances (grey) between $A$ = 150 and $A$ = 180 for comparison. The detached lower panels reveal the relative abundance change.}
\label{fig:abundA}
\end{figure*}

\begin{figure}[t]
\centering
\includegraphics[scale=0.4]{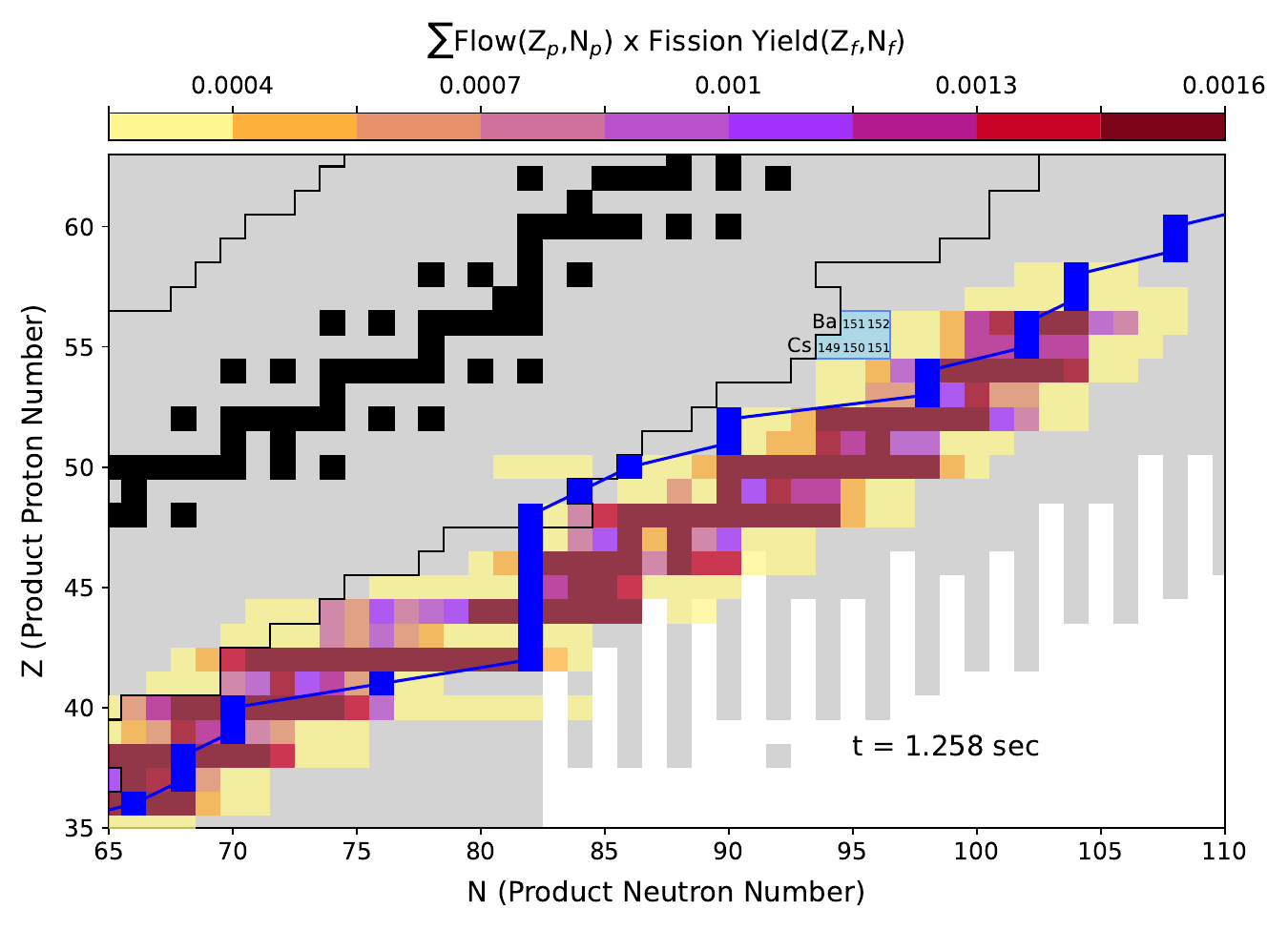}\\
\caption{Fission fragment deposition shown via fission yield weighted by flow (abundance x rate) for the fission cycling case (b) in Fig.~\ref{fig:abundA} at an instance in time right before fragments enter the light blue region probed by TITAN. The deposition strength is coded by the top color bar. The stable species are shown in black, and the black contour encloses nuclei with experimentally known masses. The FRDM2012 neutron dripline is shown in light grey. The dark blue line and boxes indicate the $r$-process path (most abundant isotopes at each proton number) at this time.}
\label{fig:NZfissionfrag}
\end{figure}

For astrophysical conditions, here we pay special attention to cases where fission deposition contributes to setting the lanthanide abundances and cases where the abundances are shaped solely from the relative stability of local isotopes. To do so, we first consider a tracer from the hydrodynamic simulation of an accretion disk from Just et al \cite{Just} which produces main $r$-process species (up through the third peak at $A\sim$195) but does not significantly populate the actinides ($Y_e\sim$0.18 for the trajectory shown here). In contrast, for a fission cycling case where late-time fission deposition can strongly influence lanthanide abundances, we consider a hot, dynamical ejecta trajectory from the simulation of Rosswog et al 2014 \cite{Rosswog}. Note that both astrophysical conditions shown in Fig.~\ref{fig:abundA} are examples of `hot' dynamics which have an extended (n,$\gamma$) $\leftrightarrow$ ($\gamma$, n) equilibrium and so are sensitive to $S_n$ trends (such as those shown in Fig~\ref{fig:NZSn}). Additionally this case is a low entropy condition with very high neutron-richness ($Y_e\sim0.01$), permitting robust fission cycling. A comparison of the $r$-process abundances with and without TITAN for both these fissioning and non-fissioning cases are shown in Fig.~\ref{fig:abundA}. For the case where actinides are not significantly populated and thus no fission products are introduced, as well as the ejecta case where the actinides are strongly populated, percent changes in the abundances near $A=149,150,151$ can reach up to $40\%$. In the case which hosts fission this is due to the influence of $^{149-151}$Cs and $^{151,152}$Ba properties on how fission fragments ultimately settle to form the final abundance pattern. This can be seen even more explicitly in Fig.~\ref{fig:NZfissionfrag} where a snapshot of the fission deposition into the lanthanide region is shown for the dynamical ejecta case by cross referencing fission flows (rate $*$ abundance) and fission yields. The $r$-process path also gives a sense of what reactions or decays nuclei are dominantly undergoing since species to the right of the path primarily undergo $\beta$-decay and those to the left of the path will mostly neutron capture. At the moment in time shown in the figure, the $r$-process abundances have yet to be finalized are are just beginning to encounter the TITAN Cs, Ba measurement region. The changes in the $S_n$, neutron capture, and $\beta$-decay introduced by TITAN's new masses thus all have an opportunity to influence the fission fragments along their journey, either by helping to set the $r$-process path during equilibrium or by influencing post-equilibrium local neutron captures, photodissociations, and $\beta$-decays. Therefore, these mass measurements push the boundaries of understanding structure in the neutron-rich lanthanide region, and correspondingly influence how fission fragments settle into place when the final abundances are forming. 

Interestingly in both the with and without fission cycling cases, the TITAN masses introduce changes that move towards smoothing over the odd-even effects in abundances. The relatively smooth isotopic pattern seen in the Solar $r$-process abundance residuals of lanthanide isotopes has long been attributed to $\beta$-decay and $\beta$-delayed neutron emission effects \cite{Cowan:2019pkx}, as well as possible deformation in rare-earth nuclei. Thus these Cs and Ba mass measurements point towards supporting the claim that neutron-rich lanthanide nuclear structure introduces a smoothing effect in rare-earth abundances.

We next explore how these mass updates affect predictions for stellar abundance trends, particularly since the region of the nuclear chart probed by these Cs, Ba measurements is populated by fission fragments during crucial times of forming the final lanthanide abundances. Stellar abundance trends have been noted to be especially revealing for lanthanide abundances, since observed ratios of lanthanides amongst themselves brought forward the idea of the `robustness' or `universality' of $r$-process abundances \cite{Beun:2007wf,Sneden2008,Goriely:2011vg,Korobkin:2012uy,deJesusMendoza-Temis:2014owk,Goriely:2015oha}: the final abundance pattern is largely insensitive to initial astrophysical conditions due to fission cycling. Further works have suggested that Ag, Pd abundances look to be correlated with Eu (a lanthanide) which can be intuitively explained if Ag (stable at $A = 107$ and $109$), Pd (stable at $A = 105$, $106$, $108$, and $110$), and Eu (long-lived at $A = 151$ and stable at $A = 153$) are co-produced through fission deposition \cite{Vassh:2019cey,Roederer:2023spd}. Following these works, the possibility for such co-production between lighter elements like Ag and lanthanides has been further explored in additional observational investigations \cite{RPAShah,RPAXD}. Since the current manuscript has probed a region of the nuclear chart that primarily impacts abundances of $A\sim148-151$, the elemental abundances that are influenced by TITAN mass measurements are Nd, Sm, and Eu. Therefore the impact on abundance ratios of relevance for co-production can be explored.

\begin{figure}[ht]
    \centering
    \includegraphics[width=0.45\textwidth]{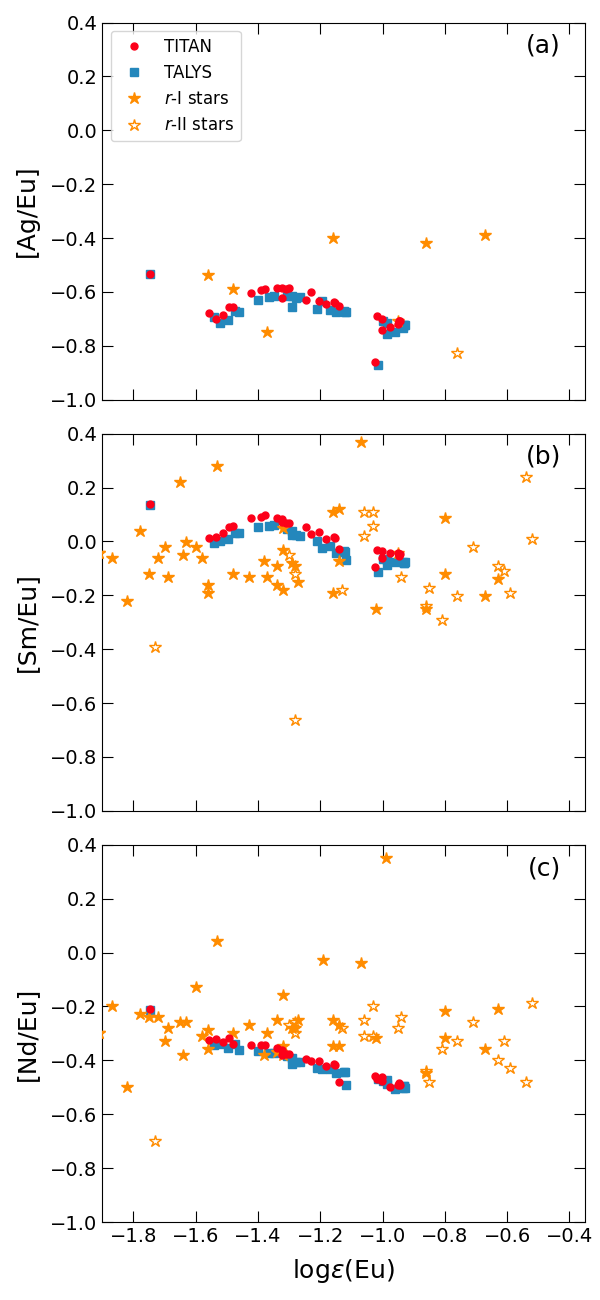}
    \caption{Predicted abundance ratios for (a) [Ag/Eu], (b) [Sm/Eu], and (c) [Nd/Eu] as a function of log$\varepsilon$(Eu )for the fission cycling cases of \cite{Rosswog}. We show in blue the results from TALYS calculations, and highlight in red the upward shift produced by the newly measured TITAN masses. Stellar abundance ratios are also included for both $r$-I and $r$-II stars.}
    \label{fig:stellarratios}
\end{figure}

To consider co-production arguments of Ag and Eu in stellar trends, in Fig.~\ref{fig:stellarratios} we first explore the abundance ratio [Ag/Eu] as a function of log$\varepsilon$(Eu) (i.e. the Eu abundance) where [A/B]=log$_{10}$(Y$_A$/Y$_B$)-log$_{10}$(Y$_A$/Y$_B$)$_{\odot}$. Here we consider the fission cycling trajectories of \cite{Rosswog} (as was considered in Fig.~\ref{fig:abundA}). In order to demonstrate that the [Ag/Eu] trend is similar to the standard trend reported for lanthanides, [Sm/Eu] is also shown. Note that for lanthanide only ratios such as [Sm/Eu] there is more stellar data to compare to than in the [Ag/Eu] case due to the greater difficulty of observing Ag in stars. Here both $r$-I (0.3 $\le$ [Eu/Fe] $\le$ +1.0 and [Ba/Eu] $<$ 0.0) and $r$-II ([Eu/Fe] $>$ +1.0 and [Ba/Eu] $<$ 0.0) stars \cite{Abohalima_2018} are considered as these are the cases which are more solidly connected to $r$-process enrichment. The flat trend of these ratios points to universality (that is, the ratio of the two is robust across stellar data), and this is visible for both [Ag/Eu] and [Sm/Eu] indicating co-production of silver and the lanthanides. Thus Fig.~\ref{fig:stellarratios} demonstrates that the new TITAN mass measurements are indeed in a region which affects the exact ratios predicted, directly affecting comparisons between theoretical calculations and stellar data, but note that the overall flat trend in the calculated ratios is preserved. Since TITAN masses also impacted the overall abundance of Nd, we also consider abundance predictions of [Nd/Eu]. In the case of Nd, the trend in predicted ratios is shifted towards the stars at higher values after the new TITAN mass data is taken into account. Thus the new TITAN mass data influences the nucleosynthesis predictions for these stellar ratios, serving as a testament to the importance of nuclear properties such as masses in interpreting stellar abundance trends. \newline

\section{Conclusions}\label{sec:conclude}
Lanthanide abundances are of interest to decipher numerous high-impact $r$-process observables such as metal-poor star abundances and kilonova. We have demonstrated that pushing the boundaries of experimentally probed species into the $A\sim150$ region of the nuclear chart for $Z=55, 56$ influences abundance predictions in both conditions that host fission and conditions that do not. We have also shown that when fission fragments are deposited into this lanthanide region, the reaction and decay properties of these exotic fragments impacts nucleosynthesis predictions. Therefore, experimental efforts to push the boundaries of experimentally probed neutron-rich properties such as these TITAN measurements, and future efforts at facilities such as ARIEL at TRIUMF and FRIB, are complementary to upcoming surveys set to gather abundances of more metal-poor stars \cite{RPAgaia1,RPAgaia2,RPAgaia3}. Interdisciplinary efforts such as these, in which experimental data is used to directly inform theoretical models and observations, will drive our field towards an understanding of the ultimate origin of elements.

\section{Acknowledgments}
The TITAN Collaboration thanks the TRIUMF Targets and Ion Source Department, which is led by A. Gottberg, for the development of the proton-to-neutron converter and their ongoing collaboration.

TITAN is funded by the Natural Sciences and Engineering Research Council of Canada (NSERC) and the National Research Council Canada (NRC). T.H.Y. and N.V. acknowledge the support of NSERC and NRC. M.R.M. was supported by the US Department of Energy through the Los Alamos National Laboratory. Los Alamos National Laboratory is operated by Triad National Security, LLC, for the National Nuclear Security Administration of U.S.\ Department of Energy (Contract No.\ 89233218CNA000001). J.B. and A.M. were supported by the German Research Foundation (DFG) under contract no. 422761894, by the German Federal Ministry for Education and Research (BMBF) under contracts no. 05P19RGFN1 and
05P21RGFN1, by Justus-Liebig-Universit¨at Gießen and
GSI under the JLU-GSI strategic Helmholtz partnership agreement.

\bibliography{CsBabib}

\end{document}